\documentclass[twocolumn,prl,showpacs]{revtex4}
\usepackage{amsmath,amssymb}
\usepackage{slashed,mathrsfs}
\usepackage{feynmp,subfigure}
\usepackage{verbatim,graphicx}
\usepackage[sub,ovp]{psfragx}
\usepackage{overpic}

\newcommand\vev[1]{\langle #1\rangle}

\begin{document}
\setlength{\unitlength}{1mm}
\title{A viable axion from gauged flavor symmetries}

\author{David Berenstein and Erik Perkins \\
{\em Department of Physics, University of California at Santa Barbara, CA 93106}}

\begin{abstract}
We consider a  string inspired non-supersymmetric extension of the standard model with gauged anomalous $U(1)$ flavor symmetries. Consistency requires the Green-Schwarz mechanism to cancel mixed anomalies.  The additional required scalars provide St\"uckelberg masses for the $Z'$ particles associated to the gauged flavor symmetry, so they decouple at low energies. Our models also include a complex scalar field $\phi$ to generate Froggatt-Nielsen mass terms for light particles giving a partial solution to the fermion mass problem. A residual approximate (anomalous) global symmetry survives at low energies. The associated pseudo-Goldstone mode is the phase of the $\phi$  scalar field, and it becomes the dominant contribution to the physical axion. An effective field theory analysis that includes neutrino masses gives a prediction for the axion decay constant.  We  find a simple model where the axion decay constant is in the center of the allowed  window.
\end{abstract}

\maketitle

\subsubsection{Introduction}
\label{S:Introduction}

The standard model fails to explain the smallness of both the QCD vacuum angle and the Yukawa couplings for light fermions. The Peccei-Quinn symmetry is a classic mechanism which addresses the smallness of the vacuum angle \cite{PQ}, and its low energy consequences are an axion whose properties can be predicted  \cite{Wilczek_78,Weinberg_78}. Such simple models are ruled out by data. To generate a viable axion the Peccei-Quinn symmetry has to be broken at scales much higher than the electroweak scale, producing an invisible axion.

By contrast the Froggatt-Nielsen mechanism~\cite{Froggatt_Nielsen_79} provides a way to obtain small masses for light fermions while maintaining $\mathcal{O}(1)$ Yukawa couplings by adding a gauged symmetry for flavor physics. Different Yukawa couplings contain different powers of the order parameter for symmetry breaking, generating hierarchies between fermion masses.

In this paper we discuss simple string-inspired models that connect these two mechanisms. In string models all symmetries should be gauged. Finding systems with approximate global symmetries usually requires fine tuning and this is considered unnatural. However, it turns out that in open string models it is possible to have approximate global gauge symmetries that are inherited from a gauge symmetry at a higher scale if the associated gauge bosons acquire a St\"uckelberg mass \cite{IQ}. We will refer to this phenomenon as the Iba\~nez-Quevedo mechanism.
If the corresponding gauge symmetry is anomalous, the Green-Schwarz anomaly cancellation mechanism will provide mass terms of that sort rendering the theory consistent.

If the symmetry we are gauging is a flavor symmetry a la Froggatt-Nielsen, then the fermion mass generation  requires an additional spontaneous symmetry breaking of the approximate global symmetry resulting in a light pseudo-Goldstone boson that can be identified with the physical axion.

 The precise string inspiration for this model comes from studying intersecting D-brane setups (see \cite{BCLS} for a review). The matter content of these models is usually presented as a quiver (or moose) diagram. We will follow these conventions. To simplify the analysis we will insist on requiring only the minimal particle content that is able to accommodate the flavor hierarchy generation mechanism at the level of effective field theory. The model we write is
 closely related to the `Minimal Quiver Standard Model' introduced in~\cite{Berenstein_Pinansky_07}. The effective field theory will have high dimension operators suppressed by some high scale $M$. For simplicity we will call such scale the string scale, although the model does not need to have a precise string theory origin. We find that if the neutrino masses are included in the analysis, we are able to determine both $M$ and the Froggatt-Nielsen (Peccei-Quinn) symmetry breaking scale, and this gives a prediction of the energy scale of the axion decay constant.

 In~\cite{Coriano_Irges_06} a similar scheme of incorporating axions into a Froggatt-Nielsen mechanism is discussed. The authors use a rather different device to break the PQ symmetry, resulting in Higgs-axion mixing and axion masses and couplings which are not strongly determined by the axion decay constant. These features are not exhibited by the physical axion appearing in our model. Also, the model in~\cite{Coriano_Irges_06} is supersymmetric, while ours is not.

Axions are a generic feature of stringy models, existing to cancel gauge anomalies via the Green-Schwarz mechanism. These axions are typically eaten by the gauge bosons whose anomalies they cure. Finding realistic axions in stringy models is not always easy, but one can not rule them out either \cite{SW}. Experimentally,  there is a narrow `axion window' which constrains the decay constants of these particles to lie between $10^{9}$ and $10^{12}$ GeV. The upper bound arises from cosmological considerations, in order to avoid overclosing the early universe. The lower bound comes from limits on energy dissipation from various stellar processes. This is reviewed in \cite{Turner_89}.
We find that in one of the models we study the axion decay constant falls exactly in this window, while other models can be ruled out because the corresponding axions are not allowed.

\subsubsection{The Models}
\label{S:Quiver}

We want a model that contains the standard model and some gauged flavor symmetries. Considerations of embedding the model into D-brane setups places various constraints on us. First, for just the standard model, the $SU(3)$ gauge color symmetry is required to be enhanced to $U(3)$ with gauged baryon number at the string scale. Furthermore, having a realistic model with minimal matter content requires an orientifold setup with two extra stacks of branes: one for the $SU(2)_W\simeq Sp(1)$ weak theory and an extra brane for a gauged $U(1)$ symmetry, so that the theory has a $U(1)\times U(1)$ gauge symmetry. Only hypercharge is a non-anomalous gauged symmetry. The other $U(1)$ symmetry
has mixed anomalies that are cancelled by the four dimensional version of the Green-Schwarz mechanism \cite{Green_Schwarz_84}. This produces a heavy $Z'$ and it is possible to
generate all couplings in the standard model at tree level, so that the only new physics is the presence of a heavy $Z'$. These considerations produce the minimal quiver model  \cite{Berenstein_Pinansky_07}.

Now, we want to make the model richer by adding a gauged flavor symmetry that distinguishes various quarks from each other. The simplest such extension requires only one more $U(1)$ gauge field, so that we have a  model with four stacks of branes. Vanishing of anomalies can be used to fit the fermions of the standard model with two up quarks and one down quark at one node, and the opposite arrangement at the other node. At this stage, we need to fit the
Higgs doublet. There are two choices: the Higgs can generate Yukawa couplings for leptons at dimension four or not, while it will for certain generate Yukawas at dimension four for some quarks. We choose these to be the heavy quarks. Given this constraint, we need to communicate the electroweak symmetry breaking to the rest of the quarks that connect to the node  where the Higgs field is absent. We do this by introducing a new complex scalar field $\phi$ between the two $U(1)$ nodes that is neutral under the standard model. This field might also be necessary to communicate electroweak symmetry breaking to the lepton sector. We will call $\phi$ the Froggatt-Nielsen (FN) scalar. These are depicted in table \ref{t:one}.
\begin{center}
\begin{table}[ht]
\begin{tabular}{|c|c|c|c|c|c|c|c|c|}
\hline
& $\bar u_R^{1,2}$ & $\bar d_R^1$ & $\bar u_R^3$ & $\bar d^{2,3}_R$ & $e_L$ & $\bar e_R$ &$\phi$& $h^{A,B}$ \\
\hline $U(1)_a$ & +1 & -1 & 0 & 0 & 0 & -1 & +1 & +1, 0\\
$U(1)_b$ & 0 & 0 & +1 & -1 & +1 & -1 & -1 & 0, +1\\
\hline
\end{tabular}
\caption{Charges under the Frogatt Nielsen symmetry. For the Higgs, we have two choices: we call the two models $A,B$}\label{t:one}
\end{table}
\end{center}

\begin{figure}[t]
\centering
\begin{overpic}[height=.25\textwidth,width=.25\textwidth]{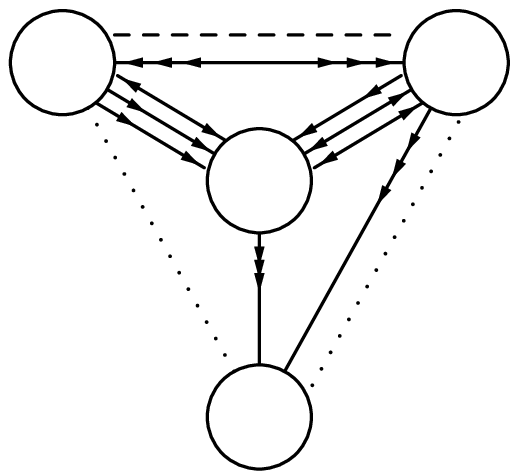}
\put(41,-2){$Sp(1)$}
\put(-9,88){$U(1)_{a}$}
\put(43,58){U(3)}
\put(91,88){$U(1)_{b}$}
\put(40,30){$q$}
\put(65,40){$l$}
\put(12,40){$h^{A}$}
\put(88,40){$h^{B}$}
\put(50,80){$\bar{e}$}
\put(50,103){$\phi$}
\end{overpic}
\vspace{3mm}
\caption{Quiver with extra scalar $\phi$\label{F:quiver}. The two choices for the Higgs doublet are labeled by the dotted lines and the model the represent $h^{A,B}$ (for the A,B models respectively). Note that if we add arrows to $\phi$ to indicate $U(1)_{ab}$ charges, they are different from those of the right handed electron.
}
\end{figure}

In the full model a $U(1)^3$ symmetry is gauged. These include $U(1)_B, U(1)_a, U(1)_b$. Hypercharge $U(1)_Y$ is a linear combination of these. Only the right handed quarks and leptons are charged under $U(1)_{ab}$, which is the set of gauge groups under which the
scalar $\phi$ is charged. We only show these in the table. Figure~\ref{F:quiver} shows the matter content in quiver form; arrow directions indicate fields in (anti)fundamental representations, e.g. up-type right-handed quarks are represented by $\rightarrow\rightarrow$ edges, while down-type quarks are represented by $\leftarrow\rightarrow$ edges. Both choices of Higgs doublet are also shown, corresponding to two different models."

Given the quantum numbers we have described, the minimal set of  couplings that generate all Yukawa couplings, including masses for left-handed neutrinos is encoded in the table \ref{t:coup}. These encode the two possible Higgs embeddings into the quiver. The high dimension operators (we will call these the Froggatt-Nielsen terms of the action) are suppressed by some high scale, $M$, which we take to be the string scale. This is a minimality assumption in order to be predictive.

\begin{center}
\begin{table}[ht]
\begin{tabular}{|c|c|}
\hline  Model A & Dimension 4 couplings \\
\hline
Quarks: & $y_{t}h^{\dagger}q\bar{t} + y_{c}h^{\dagger} q\bar{c} + y_{b}h q\bar{b}$ \\
Leptons: & $y_{e}h l\bar{e}$ \\
\hline & High dimension couplings\\
\hline
Quarks: & $\frac{y_{u}}{M}\phi h^{\dagger} q\bar{u} + \frac{y_{s}}{M}\phi^{\dagger} h q\bar{s} + \frac{y_{d}}{M}\phi^{\dagger} h q\bar{d}$ \\
Leptons:  & $ \frac{y_{\nu}}{M^{3}}\phi^{2}(h^{\dagger} l)^{2}$ \\
\hline
\hline
\hline Model B & Dimension 4 couplings \\
\hline
Quarks: & $y_{t}h^{\dagger} q\bar{t} + y_{s}h q\bar{s} + y_{b}h q\bar{b}$\\
Leptons:& $-$\\
\hline & High dimension couplings\\
\hline
Quarks: & $\frac{y_{u}}{M}\phi^{\dagger} h^{\dagger} q\bar{u} + \frac{y_{d}}{M}\phi h  q\bar{d} + \frac{y_{c}}{M}\phi^{\dagger} h^{\dagger} q\bar{c}$\\
Leptons: &$\frac{y_{e}}{M}\phi h l\bar{e} + \frac{y_{\nu}}{M}(h^{\dagger} l)^{2}$\\
\hline
\end{tabular}
\caption{List of allowed couplings}\label{t:coup}
\end{table}
\end{center}

Fields which obtain masses through Froggatt-Nielsen couplings have their masses suppressed by powers of $M$, so we interpret these as the light fields. Note that we have included the charm as a `light' quark and the strange as a `heavy quark' in model B; model A reflects the usual quark mass hierarchy. In model B, the masses of the lepton fields are suppressed, and are in the same range as those of the light quarks, which roughly mimics the observed mass hierarchy.

Following the usual Froggatt-Nielsen approach, we assume that all scalar-fermion couplings are of order 1. This is certainly true for the top quark in the standard model. The Higgs field $h$ breaks electroweak symmetry at the usual scale $v=246$ GeV. We can get an order of magnitude estimate on both $\langle\phi\rangle $ and $M$; we use the masses of the heaviest quarks and leptons with couplings of order one, and assuming that the heaviest neutrino saturates the upper bound $m_{\nu_{\tau}} \leq 2$ eV on neutrino masses, we find that

\begin{align}
\textrm{Model A:}\qquad & \langle\phi\rangle\simeq \frac{m_{s}^{3}}{m_{t}m_{\nu_{\tau}}} \simeq 5\times 10^{3}\,\textrm{GeV}\label{eq:f_1}\\
&  M \simeq \frac{m_{s}^{2}}{m_{\nu_{\tau}}} \simeq 3\times 10^{6}\,\textrm{GeV} \\
\textrm{Model B:}\qquad & \langle\phi\rangle \simeq \frac{m_{t}m_{c}}{m_{\nu_{\tau}}} \simeq  10^{11}\,\textrm{GeV},\label{eq:f_2} \\& M \simeq \frac{m_{t}^{2}}{m_{\nu_{\tau}}} \simeq 10^{13}\,\textrm{GeV}
\end{align}

Notice that we have been able to predict all relevant scales for the model from known data. The reason for this is that we have used the neutrino masses to find the scale of new physics, since it is the one parameter in the standard model that is sensitive to the high scale physics. If in a given model the neutrino masses have a different origin than the sting scale or the Frogatt-Nielsen scale, the above constraints change and tend to push the string scale higher.
At this stage we should ask what is the physical implication of these deductions for low energy physics. Before we do
that, we need to examine the effective field theory more carefully.
 
\subsubsection{Anomalies and the physical axion}
\label{S:Anomalies}

The $U(1)_{a,b,c}$ charges $Q_{a,b,c}$ have mixed $U(1)SU(2)^{2}$ and $U(1)SU(3)^{2}$ anomalies. One linear combination of these charges is anomaly free, and there is some freedom to define the two other, anomalous combinations. We take the anomalous $U(1)$ charges to be orthogonal to the anomaly-free charge. The linear combination
$
Q_{Y} = \frac{1}{6}\left(Q_{c} - 3Q_{a} - 3Q_{b}\right)
$
is anomaly-free, and corresponds to the electroweak hypercharge. This charge couples to the anomaly-free gauge field $Y_{\mu} = P_{ua}A_{\mu} + P_{ub}B_{\mu} + P_{uc}C_{\mu}$, for some mixing parameters $P_{ij}$. 

The other two $U(1)$ gauge fields are anomalous and we will call them $\tilde Z_\mu^{1,2}$. Their anomaly is cancelled via the Green-Schwarz mechanism. This requires the introduction of additional pseudoscalar fields $s^{1,2}$ with axion-like couplings to the Standard model gauge fields.
These pseudoscalar fields transform non-linearly under the anomalous $U(1)^2$ gauge transformations. Their gauge invariant kinetic term is of the form
\begin{equation}
\frac 12 K_{ij} (\partial_\mu s^i - \tilde Z_\mu^i)(\partial_\mu s^j - \tilde Z_\mu^j)
\end{equation}
and the scale of $K$ is close to $M^2$.
Since the scalars transform non-linearly, we can choose the unitary gauge where $s^{1,2}$ are constant. In this gauge
the $\tilde Z$ are clearly seen to acquire a  St\"uckelberg mass of order $M$, so they can be integrated out of the low energy effective theory. At tree level, this process produces only current-current interactions for the $U(1)^2$ gauge fields that have been integrated out, but these currents are invariant under the anomalous $U(1)$'s. Starting from this simple action, the low energy effective theory seems to have a $U(1)^2$ symmetry in the lagrangian, but in this case this symmetry is anomalous. This is a particular realization of the Iba\~nez-Quevedo mechanism \cite{IQ}. This mechanism is usually used to guarantee a long lifetime for the proton, but it serves generically to produce approximate global symmetries in the infrared effective theory. Basically, for low energy questions one can pretend that the gauged  symmetries with St\"uckelberg masses are just global symmetries of the field theory that are broken only non-perturbatively.

One of these symmetries is baryon number (this is used to prevent proton decay), while the other symmetry is the Froggatt-Nielsen symmetry of our lagrangian. The field $\phi$ is then a scalar field that
breaks the Froggatt-Nielsen symmetry, and we get an additional Goldstone boson for this symmetry breaking.
Notice that if the $U(1)_{FN}$ gauge symmetry were non-anomalous, this Goldstone boson would have been eaten by the Froggatt-Nielsen gauge field and there would be no low energy remnant. This is because having a St\"uckelberg mass for the gauge field would not be required by anomaly cancellation. Also, since the symmetry that is broken is anomalous in the low energy effective theory, the symmetry is not a true symmetry of the lagrangian anyhow and non-perturbative effects will lift the symmetry producing a potential for the would-be Goldstone boson. Such terms can be traced to the original lagrangian and they require
corrections to the potential of the form
$
V( \phi \exp( i s),\phi^* \exp(-i s) )
$, as described in \cite{Coriano_Irges_06}. These terms are not allowed by perturbative computations in string theory
so they can only be generated by D-brane instantons (see \cite{Cetal} for a review of instanton computations). This means that the would be Goldstone boson is very light and gives rise to a light invisible axion \cite{Kim_79,ZDFS}. If we take the parametrization $\phi=
r \exp(i\theta)$, then $\theta$ is the phase of $\phi$ that has been identified with an axion field at intermediate energy.

In general, a more careful analysis of the lagrangian shows that the field $\theta$ will mix with the $s^i$ fields and that the vev of $\phi$ will also contribute to the mass term of the $\tilde Z$ fields. If the $\tilde Z$ fields have a bare mass generated at the string scale, the new contribution would be a small correction at the Froggatt-Nielsen scale, so the
correction from mixing of scalars is small and $\theta$ plays the role of the intermediate energy axion.
This is what we expect in the limit of a high string scale - to a good approximation the physical axion is just the phase of $\phi$, and the axion decay constant is simply $f_a= \vev{\phi}$. This is similar to the \cite{ZDFS} model. However, the couplings to quarks and photons are different.

Since the Higgs is also charged under the Froggatt-Nielsen symmetry, when the electroweak theory is broken, there is in principle a further mixing between $\phi$ and the neutral would-be Goldstone boson from the Higgs doublet. However, at the level of the effective action at the weak scale no such terms seem to appear in the effective potential. Any such mixing would be produced from mixing with the high scale $\tilde Z'$ and is very suppressed, by $v^2/M^2$, so it can be safely ignored.

Returning to the estimates for $\vev\phi\sim f_a$ of equations (\ref{eq:f_1},\ref{eq:f_2}), we see that the axion decay constant for model B falls squarely within the nominal axion window, while that for model A falls several orders of magnitude short of the lower sill of the axion window. Model B would have been preferred anyhow because in that model the leptons have masses comparable to light quarks without fine tuning.

For completeness, we should discuss interactions of the axion with matter. There is a standard procedure for obtaining these interactions, discussed in detail in~\cite{Georgi_et_al_86}. We need to work in a basis where we have rotated the axion away from the FN terms using a PQ transformation. In doing so, we pick up derivative couplings of the axion to matter fields, as well as PQ anomaly terms for all of the standard model gauge fields. By a similar procedure~\cite{Georgi_et_al_86,Kaplan_85}, one also obtains derivative couplings of the  axion to baryons.

This gives a derivative pseudovector coupling $g_{aNN}(\partial_{\mu}a/f_{a})\bar{N}\gamma^{\mu}\gamma_{5}N$ to nucleons.  The relevant couplings for astrophysical bounds on $f_{a}$ are $g_{app} = -0.54/f_{a}$ GeV$^{-1}$, $g_{ann} = 0.61/f_{a}$ GeV$^{-1}$ for model B. These were derived, in the notation of~\cite{Kaplan_85}, using $F=0.47,\,D=0.81,\,S=0.13$ \cite{PDG} (see also the review \cite{Kim_C_08}).
The next interaction to address is photon-axion coupling. We find  that
$C_{a\gamma\gamma}\simeq -1.2$ is of order one. Hence the axion couplings to matter are ordinary (or order one) and there is no accidental cancellation that can change our estimates of the viability of the axion model. Similar results can be found for model A.
A more detailed study of these models will appear elsewhere \cite{BP}.

To conclude, we find that some string inspired minimal models for flavor are able to predict axions within the allowed window by assuming that the scale of new physics (taken as the string scale) is closely related to the scale at which the neutrino masses are generated. The models also predict an intermediate string scale that can be rather light. If the string scale is too light, the axion falls outside the allowed window and such models are essentially ruled out. It would be interesting to study further how such models can be embedded in supersymmetric setups so that the hierarchy problem can also be addressed.

{\em Acknowledgements:} We would like to thank M. Srednicki for discussions. Work supported in part by DOE under grant DE-FG02-91ER40618.


\end{document}